# The effect of dust composition and shape on radiation-pressure forces and blowout sizes of particles in debris disks

Jessica A. Arnold[1], Alycia J. Weinberger[1], Gorden Videen[2], Evgenij S. Zubko[3]

[1]Department of Terrestrial Magnetism, Carnegie Institution for Science, 5421 Broad Branch Rd. Washington, DC 20015, USA (*jarnold@carnegiescience.edu*), [2]Space Science Institute, 4750 Walnut Street, Boulder Suite 205, CO 80301, USA, [3]School of Natural Sciences, Far Eastern Federal University, 8 Sukhanova Street, Vladivostok 690950, Russia

**Abstract**

The light scattered from dust grains in debris disks is typically modeled as compact spheres using Lorenz-Mie theory or as porous spheres by incorporating an effective medium theory. In this work we examine the effect of incorporating a more realistic particle morphology on estimated radiation-pressure blowout sizes. To calculate the scattering and absorption cross sections of irregularly shaped dust grains, we use the discrete dipole approximation. These cross sections are necessary to calculate the $\beta$-ratio, which determines whether dust grains can remain gravitationally bound to their star. We calculate blowout sizes for a range of stellar spectral types corresponding with stars known to host debris disks. As with compact spheres, more luminous stars blow out larger irregularly shaped dust grains. We also find that dust grain composition influences blowout size such that absorptive grains are more readily removed from the disk. Moreover, the difference between blowout sizes calculated assuming spherical particles versus particle morphologies more representative of real dust particles is compositionally dependent as well, with blowout size estimates diverging further for transparent grains. We find that the blowout sizes calculated have a strong dependence on the particle model used, with differences in the blowout size calculated being as large as an order of magnitude for particles of similar porosities.

## 1. Introduction

Several hundred nearby main sequence stars are known to host debris disks (e.g. Wyatt 2008; Montesinos et al. 2016) containing micron- to millimeter-sized dust (e.g. Rodigas et al. 2015; Hughes, Duchene & Matthews in press). At visible and near-infrared wavelengths, dust within debris disks is detected via light from the host star scattered by these dust grains. While larger dust grains are gravitationally bound to their star, smaller grains are subject to radiation forces that can cause them to be removed from the disk. These forces are Poynting-Robertson (P-R) drag, which is a tangential force, causing dust to spiral inward, and radiation pressure, which is radial and causes dust to be blown out of the disk (Burns et al. 1979). Typically P-R drag is not considered because in order for the P-R dust removal timescale to be shorter than the collisional timescale, the disk would need to have so little mass as to be undetectable (Wyatt 2005, 2011). Moreover, when grains are sufficiently small, the radiation pressure is much stronger than P-R drag and these grains will be blown out of the disk. Hence, to interpret scattered light observations of debris disks, the minimum grain size is usually assumed to be approximately the radiation pressure blowout size (e.g. Krivov et el. 2006; Pawellek and Krivov et al. 2014).

One convenient parameter for examining the blowout size is the ratio of radiation pressure to gravitational forces ($\beta = F_{rad}/F_{grav}$) acting on a dust grain, which depends on grain composition, size, and structure. Typically, the $\beta$-ratio is calculated using the assumption of compact, spherical particles or porous, spherical particles accounting for porosity via an effective medium theory (EMT) approximation such as the Maxwell-Garnett or Bruggeman mixing rule (e.g. Mukai 1992; Lebreton et al. 2012). However, real circumstellar dust grains are expected to take on complex, porous structures similar to interstellar (ISD), interplanetary (IPD), and cometary dust particles.

Although debris-disk dust grains cannot be studied in situ, we can gain some insight from dust within our own solar system. IDPs, thought to be derived from either asteroid or cometary materials, with a small presolar grain component (Brownlee, Joswiak, and Matrajt 2012), have a wide variety of shapes (e.g. Brownlee 1985). Comet grains from 81 P/Wild 2 were shown to have diverse shapes and densities from impacts into the collector surface returned by the Stardust mission (Hörz et al. 2006). Moreover, MIDAS, an atomic force microscope instrument on Rosetta has shown that cometary dust grains within comet 67P/Churyumov-Gerasimenko have similar shape and composition to IDPs (Bentley et al. 2016). Telescopic data also provides evidence for porous, irregularly shaped cosmic dust grains. Irregularly shaped particles have been shown to produce a better match to the interstellar silicate extinction spectrum (Min et al. 2006) and can be used to model photo-polarimetric data of comets (Mukai and Mukai 1990; Zubko et al. 2014, 2015).

Calculations of radiation pressure-to-gravity ratios for porous, irregular dust grains have been carried out for a handful of cases (Kimura and Mann 1999; Kimura et al. 2002; Köhler et al. 2007; Silsbee and Draine 2016) using the discrete dipole approximation (DDA) method (e.g. Draine and Flatau 1994). However, due to computational considerations, these studies focused on submicron particles that only require a small number of dipoles (N<=2048), but are well below the blowout size of most debris disk systems. Non-spherical dust morphologies previously considered using the DDA method include sub-micron ballistic particle-cluster aggregates (BPCA) and ballistic cluster-cluster aggregates (BCCA) (Kimura and Mann 1999; Kimura et al. 2002; Köhler et al. 2007; Silsbee and Draine 2016). Kirchschlager and Wolf (2013) included larger particle sizes up to 10 times the wavelength, but considered a simplified geometry generated by random removal of dipoles from a sphere. While this simplified approach allows the calculation to converge more rapidly, the calculated scattering and absorption cross-sections display Mie resonance when the grain size is similar to the wavelength. This oscillatory behavior occurs due to multiple internal reflections within the spherical particle that cause constructive interference (van de Hulst 1981).

Here we compare calculations of the $\beta$-ratio using three different combinations of scattering approximations and grain models, compact spheres using the Lorenz-Mie theory, porous spheres using Lorenz-Mie theory and the Bruggeman EMT, and irregularly shaped agglomerated debris particles using the DDA. We consider micron-sized grains of varying composition surrounding stars of different stellar spectral types. The agglomerated debris particle shapes and DDA implementation used to generate scattering and absorption efficiencies are similar to those used by Zubko et al. (2005). Some examples of these particles are shown in **Figure 1**. Stellar properties were chosen to correspond to stars known to host debris disks and cover the range of main sequence spectral types: HR 4796A, β Pictoris, HD 32297, HD 181327, BD +20°307, HD 61005, ε Eridani, and AU Microscopii. We also include calculations for the solar system.

## 2. Methods

### *2.1 Dust grain blowout size*

When the force of radiation pressure exceeds that of gravity the dust grain becomes unbound from the system on an orbital timescale. The ratio of these forces is often expressed as

$$\beta = F_{rad}/F_{grav} = \frac{\sigma\langle Q_{pr}\rangle L_*}{4\pi GmcM_*}, \tag{1}$$

where σ is the grain cross-sectional area, $\langle Q_{pr}\rangle$ is the average radiation pressure efficiency over the stellar spectrum ($F_\lambda$), $G$ is the gravitational constant, $m$ is the grain mass, $c$ is the speed of light, and $L_*$ and $M_*$ refer to the luminosity and mass of the host star, respectively (Burns et al. 1979, Krivov et al. 2006). The radiation pressure efficiency is defined as

$$Q_{pr} = Q_{abs} + (1 - g)Q_{sca}, \tag{2}$$

where $Q_{abs}$ and $Q_{sca}$ are the absorption and scattering efficiencies respectively, and $g$ is the cosine asymmetry parameter. The efficiencies in $Q_{pr}$ depend on dust grain composition (i.e., the complex index of refraction as a function of wavelength, denoted by *n*, and *k* for the real and imaginary parts, respectively), size ($a$), shape, and porosity ($\mathcal{P}$). Assuming a grain with a spherical cross-section and restating in terms of solid density ($\rho$) and porosity ($\mathcal{P}$),

$$\beta = \frac{3\langle Q_{pr}(n,k,a,\mathcal{P})\rangle L_*}{16\pi Gc(1-\mathcal{P})\rho a M_*}. \tag{3}$$

Porosity is defined as the fraction of a circumscribing sphere that is taken up by void space,

$$\mathcal{P} = {V_{void}}/{V_{total}} = 1 - {V_{solid}}/{V_{total}}. \tag{4}$$

The solid density is simply the sum of the density of each component multiplied by the fraction of the solid volume each material occupies so that, $\rho = \sum_j v_j \rho_j$, where $v_j = V_j/V_{solid}$. It should be noted that using a circumscribing sphere to define porosity will not be appropriate in all cases. If a particle is a highly elongated ellipsoid for example, the calculated porosity will end up spuriously high. In the case of agglomerated debris particles the geometry of the particle is determined by filling nodes in a sphere (as described in section 2.3), so this is an appropriate comparison volume.

As can be seen from Eq. 3, when all other variables are fixed decreasing particle radius ($a$) will result in increasing $\beta$, and if this ratio exceeds 0.5, the grain gets pushed into a hyperbolic orbit and leaves the system (Krivov et al 2006). This cut-off assumes that the grain is released from a parent body in a low-eccentricity orbit. However, as discussed in Kral et al. 2013, grains close to β=0.5 with low initial eccentricity will quickly be driven to highly eccentric or unbound orbits by radiation pressure. Hence, for a given grain composition and porosity, this $\beta$ parameter cut-off can be used as a very close approximation of the effective radius below which grains will not stay in the system. The radius at which this threshold is reached ($a_{BO}$) is known as the blowout size, and re-arranging the above yields

$$\left(\frac{a_{BO}}{\mu m}\right) = 1.152\langle Q_{pr}\rangle\left(\frac{L_*}{L_\odot}\right)\left(\frac{M_*}{M_\odot}\right)^{-1}\left(\frac{(1-\mathcal{P})\rho}{g/cc}\right)^{-1}. \tag{5}$$

### *2.2 Light-scattering calculations*

To represent debris-disk dust grain compositions, we use the wavelength-dependent optical constants, i.e. the real and imaginary components of the index of refraction ($m = n + i\kappa$) of astronomical silicate (Draine and Lee 1984; Draine 2003b,c) with the chemical formula $MgFeSiO_4$, amorphous carbon (Zubko 1996), and water ice (Henning and Stognienko 1996, online database). Typical mass opacities for silicate and amorphous carbon are given in **Figure 1**. Light scattering calculations depend on the ratio between the particle radius and wavelength, which is expressed in terms of a quantity known as the size parameter, $x = 2\pi a/\lambda$. Mass opacity is related to the wavelength-dependent optical constants as follows, $\kappa(x, n, k) = \frac{4Q_{abs}(x,n,k)}{3\rho a}$. The specific gravity of each of these are 3.5, 2.2, and 0.9, respectively (Draine 2003a). For each grain composition, we compare three different light-scattering models and associated grain shapes: 1) compact spheres where $\langle Q_{pr}\rangle$ is calculated using Lorenz-Mie theory, 2) porous spheres where the Lorenz-Mie theory $\langle Q_{pr}\rangle$ is calculated using optical constants modified by the Bruggeman mixing rule (Bruggeman 1935) and 3) irregular grains where $\langle Q_{pr}\rangle$ is calculated via the DDA method. Because the algorithm used to generate irregular agglomerated debris particle dust grain shapes yields roughly 70-80% porosity with a mean value of 76.4% (Zubko et al. 2015), we compare the results from the agglomerated debris particles to the Lorenz-Mie-Bruggeman calculations with 76.4% porosity.

#### *2.2.1 Porous spheres*

The Bruggeman mixing rule approximates a compositionally heterogeneous particle by assigning an effective complex refractive index, $m_{eff}$. The effective complex refractive index is found by solving for the root of the following equation,

$$\sum_j \mathbb{V}_j \frac{\varepsilon_j - \varepsilon_{eff}}{\varepsilon_j + 2\varepsilon_{eff}} = 0, \quad (6)$$

where $\mathbb{V}_j$ is the volume fraction of each material $j$, with a permeability ($\varepsilon_j = m_j^2$) that combine to give a complex dielectric constant, $\varepsilon_{eff} = m_{eff}^2$. To obtain an effective index that does not include porosity, $\mathbb{V}_j = v_j$, i.e., the volume fraction of solid material. To model a porous material, $\mathbb{V}_j$ is the fraction of the total volume including pore space in which one of the materials is vacuum with $\mathbb{V}_j = \mathcal{P}$ and $m = 1 + i0$. The latter model is the one that we use for porous spheres. **Figure 1** compares the mass opacity of solid spheres with 76.4% porosity spheres for a few different compositions.

#### *2.2.2 Agglomerated debris particles*

We use the implementation of the DDA method from Zubko et al. (2010) to calculate the light-scattering properties of irregularly shaped agglomerated dust grains. Within DDA, particle shapes are generated by placing electric dipoles within a 3D cubic lattice of points. Each dipole has a specified complex index of refraction. When reporting the size parameters for these particles we use $a$ as the radius of the circumscribed sphere of the agglomerated debris particle. The spacing between dipoles is constrained by both the grain size and the wavelength and must meet the condition $|m|\hbar d < 1.0$ (Zubko et al. 2010, Draine and Flatau 1994), where $d$ is the lattice spacing and $\hbar = {}^{2\pi}/_{\lambda}$ is the wavenumber. This criterion is used to calculate the number of dipoles appropriate for a given size parameter, $N_{dipoles}(d) = \frac{4}{3}\pi \frac{a^3}{d^3}$, so since $x = \hbar a$, $N_{dipoles} > \frac{4}{3}\pi(x|m|)^3$. We use this as a rough guide for the minimum lattice length as the actual number of dipoles will depend on the amount of void space in the randomly generated particle.

As the size parameter increases, so do the number of dipoles and therefore the necessary computation time. For example, a particle with a size parameter of 10 (~$4\text{x}10^4$ dipoles), the average scattering properties can be calculated in approximately 5 days using a single 2.2 GHz cpu core; whereas, one of our test cases with a size parameter of 48 (~$2\text{x}10^6$ dipoles) took approximately 80 days using 20, 2.5 GHz CPU cores.

The agglomerated debris particle shape (examples, **Figure 2**) is generated by confining the dipole lattice points to a sphere and filling the interior points as follows. The upper 0.5% of the sphere diameter is designated as a surface layer. Below the surface layer, 21 points are randomly chosen as seeds for material while another 20 are randomly chosen as seeds for void space. The allocation of void space versus material determines the porosity of the grain. Within the surface layer, a further 100 points are selected as seed cells of void space. Each non-seed lattice point is assigned the properties of the nearest seed particle, whether it is a material or empty space. This generates randomly shaped solid blocks of material (aka monomers) interspersed with pore spaces of a similar size. The average porosity for these particles is 76.4% (Zubko et al. 2015) in comparison to a solid sphere. To get an estimate of the error that might be associated with using a sphere for comparison rather than a more exact shape, we used De-launay triangulation to produce faceted enclosing shapes for 6 of the particles. Using these more exact circumscribing shapes, the average porosity was slightly lower at 70%.

Scattering properties are averaged over at least 500 such randomly generated particles placed in random orientations relative to the direction of incident light. More particles are added as necessary until the addition of new particles changes the standard deviation of the degree of linear polarization by less than 1% for each scattering angle in order to ensure that the results are not orientation dependent.

**Table 1** gives the size parameters, lattice size, and indices of refraction of the agglomerated debris particles. The range of particle sizes and wavelengths that can be used for agglomerated debris particle $\beta$-ratio calculations is limited by the size parameter values that have been calculated to date. We choose which particle sizes to calculate $\beta$-ratios based on the need to cover the peak wavelengths of the stellar spectrum and the available size parameters. Both size parameters and indices of refraction are a function of wavelength, and we choose the closest available values for each wavelength.

### *2.3 Stellar properties*

We consider 8 debris disk systems, in addition to the solar system, that have been spatially resolved; their properties are given in **Table 2.** These properties are used to generate a synthetic stellar spectrum over which radiation pressure efficiency is averaged, $\langle Q_{pr}\rangle = \int Q_{pr,\lambda} F_\lambda d\lambda / \int F_\lambda d\lambda$. The minimum wavelength ranged from 0.1 μm for the hottest star HR 4796 to 0.4 μm for the coolest star AU Mic. The maximum wavelength ranged from 1 μm for HR 4796 to 2.5 μm for AU Mic. For each star we considered 20 wavelengths. Adding more wavelengths increases the computation time without changing the results. We calculated the stellar luminosities by fitting Kurucz models to the visible and near-infrared photometry of the stars, converting to absolute flux densities with the Gaia DR2 distances, and integrating under the model to find the luminosity. One of these, BD +20°307 is a spectroscopic binary system. Both stars of the binary BD +20°307 are of equal mass and approximately the same spectral type. Hence, we use $M_* = 0.9\ M_\odot$ when computing the stellar spectrum, $M_* = 1.8\ M_\odot$ for the total mass of the system when computing $\beta$, and use the total luminosity of the binary in Eq 3.

## 3. Results

### *3.1 Blowout size as a function of absorption and stellar spectral type*

**Figures 3-11** show $\beta$-ratio vs. grain size ($a$) for each of the three light-scattering models described in Section 2.1 calculated for eight nearby stars with debris disks as well as the solar system. **Figures 3A-11A** present calculations for single-composition grains including amorphous carbon, astronomical silicate, and water ice; and **Figures 3B-11B** present calculations for grains of mixed composition. Particles with $\beta$-ratio less than 0.5 are sub blowout, meaning that absent other processes they can remain in the disk. For all three light-scattering models tested, increased stellar luminosity increases the radiation pressure exerted on grains of a given size and composition, which in turn raises the blowout sizes ($a_{min}$). Additionally, the maximum $\beta$-ratio ($\beta_{max}$) shifts to smaller grain size as luminosity increases. Moreover, both $a_{min}$ and $\beta_{max}$ occur at different grain sizes for the same composition depending on the light-scattering model used. They also depend on the fraction of highly absorbing materials, in this cases amorphous carbon.

Blowout sizes (**Table 3**) calculated for a given composition assuming compact spheres are smaller than those calculated assuming porous spheres or agglomerated debris particles. The discrepancy in $\beta$ between the porous spheres and agglomerated debris particles having the same porosity depends on the grain composition. Near $\beta_{max}$, the irregular, agglomerated debris particles behave similarly to porous spheres when composed of amorphous carbon, but not astronomical silicate. However, when comparing blowout sizes, it is the silicate particles that have more similar values between the two morphologies. For carbon grains, the agglomerated debris particles have significantly smaller blowout sizes than the porous spheres (3.1 μm vs. 4.9 μm in the case of HD 181327). The two morphologies have comparable values for astronomical silicate grains for the three most luminous stars considered. However, in the remaining cases, the silicate grains are either sub-blowout for both geometries (3 systems) or the agglomerated debris particles have a blowout size, while the porous spheres are sub-blowout (3 systems).

DDA calculations of the large grain sizes needed to approach the blowout size of A-type stars such as HR 4976A or HD 32297 would require prohibitively long computation times. The largest size parameter that we calculated, $x = 48$, took roughly two months to compute. However, in Table 4 we have provided two sets of fits to the computed points for each disk. The first fit is a 6-parameter Moffat function for the size range of the computed points. These are provided so that the $\beta$ values can be used by other authors without having to do the time-consuming DDA calculations and are not meant to be used outside of the grain size range of the points given in **Figures 3-10**. The second fit is a linear extrapolation, which can be used to estimate the agglomerate blowout sizes for the more luminous stars and extrapolate the $\beta$-ratio plot past the grain size of $a = 100\ \mu m$. For HD 181327 through AU Mic, the linear fits were obtained from the four largest grain sizes, as these points are well past the $\beta_{max}$ peak. In the cases of HR 4796, β Pic, and HD 32297, the slope was constrained to be that of the linear fit for AU Mic, the lowest luminosity case. This slope value is able to produce a good fit for these three cases and these two approaches give very similar values for HD 181327, hence why we picked this as the transition point.

*3.2 Blowout size of mixtures*

Actual debris disk dust grains may be composed of intimate mixtures of silicate, carbon and ice. We use the Bruggeman mixing rule to calculate an effective index of refraction for the solid portion of the irregular dust grains as was done in Videen et al. (2015). Videen et al. (2015) showed such an EMT approximation works for mixtures of two solid components with a wide range of differences in absorption between the components. We use this effective index of refraction of solids for the agglomerated debris particles. However the effectiveness of EMT diminishes when one of the components is a vacuum as is the case in the porous spheres. In **Figures 3B-10B** we present results for two 2-component mixtures and one 3-component mixture. These are, in terms of volume percentage: 1) 95% amorphous carbon and 5% water ice; 2) 95% astronomical silicate and 5% amorphous carbon; and 3) 33% astronomical silicate, 33% amorphous carbon, and 34% water ice.

For a given composition, the overall trends for $a_{min}$ and $\beta_{max}$ with luminosity agree with the single-component cases, with both $a_{min}$ and the position of $\beta_{max}$ decreasing in grain size with luminosity. We find that the effectiveness of using the Bruggeman mixing rule to represent porosity depends on the amount of highly absorbing material, in this case amorphous carbon. This is consistent with our results for amorphous carbon and silicate endmembers. Similar to the pure carbon case, the largest discrepancy in blowout sizes occurs for the 95% carbon 5% ice mixture (e.g. for HR 4796A the agglomerated debris blowout size is 11 μm and the Lorenz-Mie-Bruggemen blowout size is 24 μm). The smallest discrepancy in blowout size between the agglomerated debris particles and the porous spheres occurs for 95% silicate 5% ice mixture (13 μm vs 15 μm for HR 4796A, respectively).

## 4. Implications

The $\beta$-ratio model presented here suggests that grain shape and porosity can have a significant effect on the inferred blowout size, and must be taken into account in such calculations. The porosity cannot be accounted for using optical constants calculated using an EMT such as the Bruggeman mixing rule. Kirchschlager and Wolf (2013) used the DDA method to calculate blowout sizes for porous, spherical astronomical silicate grains with a morphology that resembles Bruggeman's model, a sphere with random dipoles removed until the desired porosity is reached. Although that work only explores porosities of up to 60%, it is still useful to compare their calculated blowout sizes for comparable stellar temperatures with our Bruggeman EMT and agglomerate DDA results. The Kirchschlager and Wolf (2013) porous sphere DDA calculations give a blowout size similar to our agglomerate DDA calculations, despite the assumed geometry being identical to that of Bruggeman EMT. Kirchschlager and Wolf (2013) computes a blowout size of ~0.7 μm for an astronomical silicate particle with 60% porosity around a sun-like 5770 K, 1 $M_{\odot}$ star; whereas this work (**Table 3 and Figure 11**) finds that for 76.4% porosity and a 5778 K, 1 $M_{\odot}$ star the agglomerated debris particles have a blowout size of 0.9 μm and the porous particles approximated by the Bruggeman mixing rule are all sub-blowout. Hence, EMT likely underestimates the β-ratio for silicates.

The ~2-3 μm minimum grain size modeled for HR 4976A by Telesco et al. (2000) using thermal infrared images at 10.8 and 18.2 μm and by Rodigas et al. (2015) using VNIR scattered light images is smaller than that implied by compact, spherical silicate grains. This mismatch is exacerbated when porosity is taken into account. Hence, either smaller, sub-blowout size particles are being continuously generated within this system, or the minimum sizes derived from telescopic data using Lorenz-Mie spheres are not accurate. A similar mismatch occurs for HD 32297. The minimum grain radius within both the inner and outer disks modeled by Donaldson et al. (2013) is close to the 1 μm blowout size of compact, spherical silicate grains. However, while the inner disk is expected to contain compact grains, the outer disk is expected to have very porous fractal-like grains. According to the models presented here, these smaller grains should be blown out of the outer disk.

Ballering et al. (2016) presented best fits for dust composition and size within the β Pictoris disk based on 5 images collected with a variety of instruments at wavelengths ranging from 0.58 μm to 870 μm. Again, while the best fits required the presence of compact, sub-micron dust, this is well below their calculated blowout size of their best fit composition, 2.7 μm. Tamura et al. (2006) modeled K-band polarimetry of β Pic using astronomical silicate and the constraint $a_{min} = 2.0\ \mu m$. This agrees with our calculation of 1.8 μm for a compact spherical grain of the same material. Tamura et al. found this grain model provided good agreement with the polarization data, but noted that mid-infrared spectral data showed that grains were likely not compact, but rather fluffy and porous. However, porous grains would likely have a much larger blowout size. They proposed that the grains may consist of optically dark dust aggregates encased in ice, which would coincidentally have similar refractive indices and blowout sizes to compact silicate.

Lebreton et al. (2012) examined several models for HD 181327, taking into account both a range of porosities and mixture compositions. The best-fit 4-component model including carbon, silicate, ice and porosity implied a blowout size 4 times that of the modeled minimum grain size. Models including no porosity or carbon were closer to the inferred minimum grain size. Our models show that blowout sizes calculated using agglomerated debris particles are smaller than those for porous spherical grains for all compositions except 95% and 100% silicate. However, none of these porous models approaches their preferred < 1 μm minimum grain size of HD 181327. (n.b. Figure 7 of Lebreton et al. 2012 is either not correct in the units given, or may have been rescaled in some way not described therein. c.f. Figure 1 of this work. However, our calculations of β agree). Stark et al. 2014 suggest that interactions with the ISM could allow the presence of enough sub-blowout grains to produce a significant contribution to the disk scattering properties.

Overall, for disks surrounding stars luminous enough to blow out micron-sized grains, the grain sizes that best model the scattered light are smaller than the blowout sizes. In contrast, minimum grain sizes based on fits to mid and thermal infrared SED's often exceed the estimated blowout size. Roccatagliata et al. (2009) performed SED analyses for disks around six sun-like stars and found that $a_{min}$ exceeded $a_{BO}$ for four of these systems. Ertel et al. (2009) compared the minimum grain size to the blowout size for the SED of HD107146 and found that $a_{min}$ is several times larger than $a_{BO}$. Ardila et al. (2004) infers a sub-micron blowout size from scattered light images for the same disk. Ertel et al. (2009) also found a discrepancy between the mass estimates for HD107146 based on scattered light vs infrared SED datasets.

Brunngräber et al. (2017) produced model disks containing spherical particles of varying porosity and then fit these simulated SEDs with compact spheres. The resulting fits for $a_{min}$ based on compact spheres differ from the true starting values by up to a factor of 2. Just as compact, spherical grains produce much different estimated blowout sizes than irregular, porous grains; scattered light models based on irregularly shaped agglomerated debris particles will likely produce different results for minimum grain sizes as well. Hence, the next step is to produce self-consistent models where the blowout size and scattered light data are modeled using the same particle morphology. Similarly, future work should test whether scattered light models based on agglomerated debris particles can fit disk color and albedo, while remaining consistent with the blowout size modeled for those same agglomerated debris particles.

Not all systems have a blowout size. This is particularly true of low luminosity stars. Moreover, for some systems and grain compositions, the $\beta$-ratio curve crosses the 0.5 threshold at two points, suggesting there are two blowout sizes. For example, this is the case for pure and 95% astronomical silicate grains within the BD +20°307 disk (**Figure 7**). Hence, for the smaller particles of β < 0.5, other forces (P-R drag and stellar wind drag) need to be examined. In the AU Mic system, most compositions are expected to be able to persist within the disk regardless of grain size. However, the results calculated using the agglomerated debris model suggest pure amorphous carbon may be pushed out of the system for a narrow range of grain sizes near $\beta_{max}$.

Our agglomerated debris models suggest different minimum grain sizes for different grain compositions, whereas models of scattered light images typically assume, or aim to fit, a single minimum size. In the HD 61005 and ε Eridani disks, pure silicate or water ice is always sub-blowout for both compact and agglomerated debris particle grains. However, the addition of 33% amorphous carbon to the grain increases the effect of radiation pressure to the point where there is a blowout size for both compact and agglomerated debris particle grains.

## 5. Conclusions

We have demonstrated the necessity of modeling particles having realistic shape and treatment of light-scattering properties for the accurate calculation of dust blowout sizes. Blowout sizes calculated from agglomerated debris particles are larger than previously inferred assuming compact spheres. We have also shown the following:

1) Blowout size estimates for a given porosity can differ significantly between agglomerated debris particles calculated with DDA and those calculated using the Lorenz-Mie theory and an EMT such as the Bruggeman mixing rule.
2) The difference between these two models for dealing with porosity and shape varies in a compositionally dependent way.
3) Using agglomerated debris particles rather than an EMT to account for porosity does not resolve the discrepancies between blowout sizes, modeled minimum grain sizes for scattered light images, and modeled minimum grain sizes based on infrared SEDs of disks around luminous stars.
4) The range of particle sizes and wavelengths that can be used for agglomerated debris particle $\beta$-ratio calculations is limited by the size parameter values that have been calculated to date. Therefore it is important to continue to expand the database of these calculations for use in future work.

Blowout-size estimates for irregular agglomerated debris particles and porous spherical particles of the same porosity are most similar for particles consisting of pure astronomical silicate. For highly absorbing, pure amorphous carbon grains, porous spheres produce much (2x) larger blowout sizes than agglomerated debris particles, while for weakly absorbing, pure silicate grains, porous spheres produce slightly smaller (~4-8%) blowout sizes than agglomerated debris particles for all but the most luminous star (HR 4796A).

Models of the total brightness and color for several disks including HR 4976A, β Pictoris, HD 32297, and HD 181327 require sub-blowout size grains to be present in the disk. Accounting for both irregular grain shape and porosity increases the blowout size, and therefore adds to this inconsistency. Although blowout sizes for carbon-rich grains are smaller for agglomerated debris particles than for porous spheres, and closer to those of compact spheres, this composition is not necessarily realistic and does not fully account for the difference. Both blowout size calculations and models of photo-spectroscopic scattered light images depend on the scattering efficiency $Q_{sca}$. However, scattered light images of disks are typically modeled using spherical particles. Hence, the next step is to compare self-consistent models for blowout size and scattered light that both more realistic particle shapes such as agglomerated debris particles to calculate $Q_{sca}$ of disk dust grains.

## Acknowledgments

Support for Program number HST-AR-14590.002-A was provided by NASA through a grant from the Space Telescope Science Institute, which is operated by the Association of Universities for Research in Astronomy, Incorporated, under NASA contract NAS5-26555. This project is supported by NASA ROSES XRP, grant NNX17AB91G. We would like to thank an anonymous referee for their thorough review and detailed comments which have improved the quality of the manuscript.

**Figures and Tables**

Figure 1: Mass opacity of compact spherical grains computed with Lorenz-Mie theory (solid lines) compared with those of porous spheres computed with the Bruggeman mixing rule (dashed lines) and the agglomerated debris particles (symbols). The grain size distribution follows the power law $dN \propto a^{-3.5} da$, with a minimum and maximum grain size of 0.5 and 2 μm respectively.

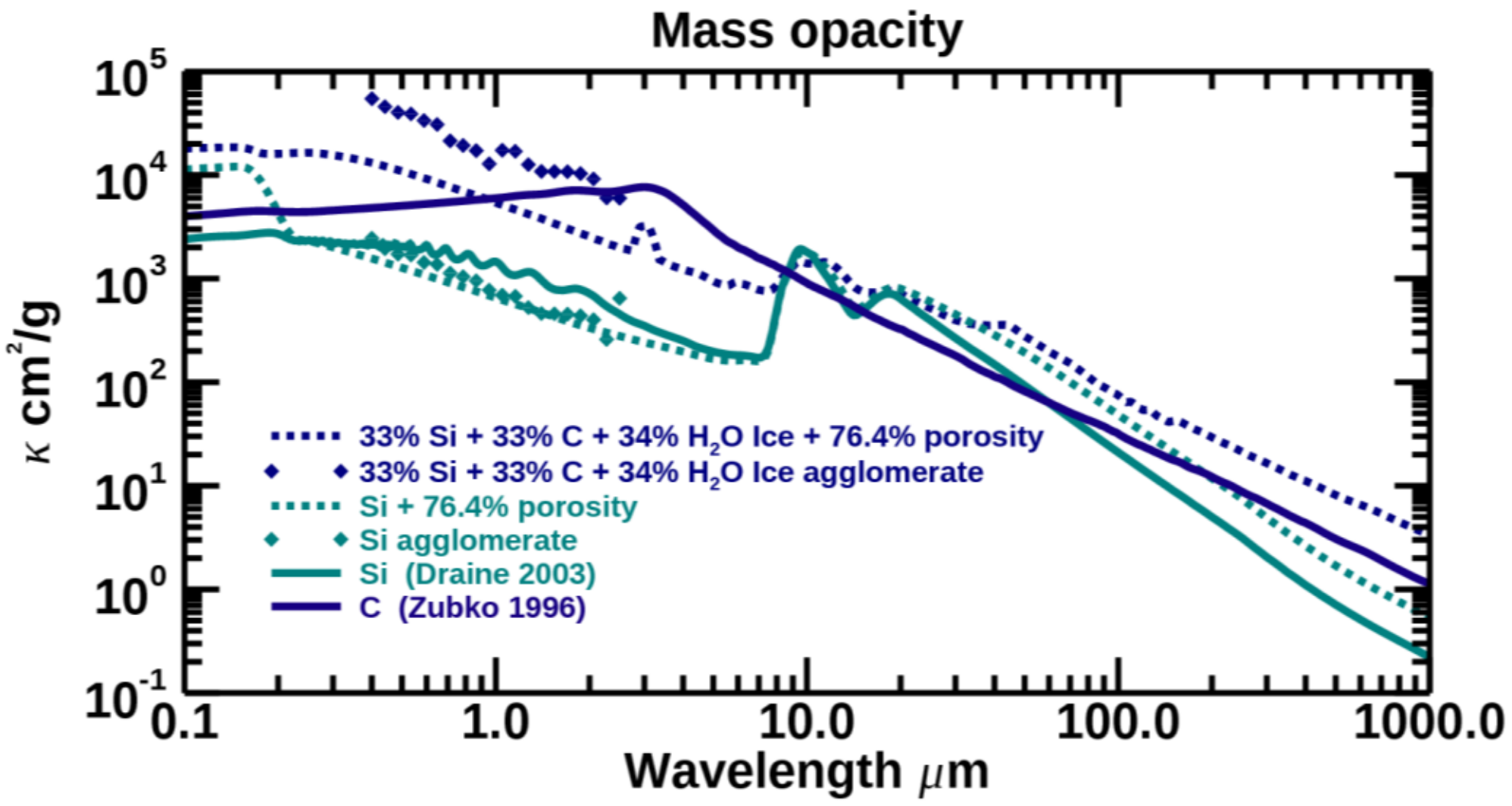

Figure 2: Six examples of agglomerated debris particles generated by the process outlined in Section 2.3. These were generated with lattice dimensions of $256^3$ dipoles, which was cut down to $64^3$ dipoles for rendering purposes. This type of particle has an average porosity of 76.4%.

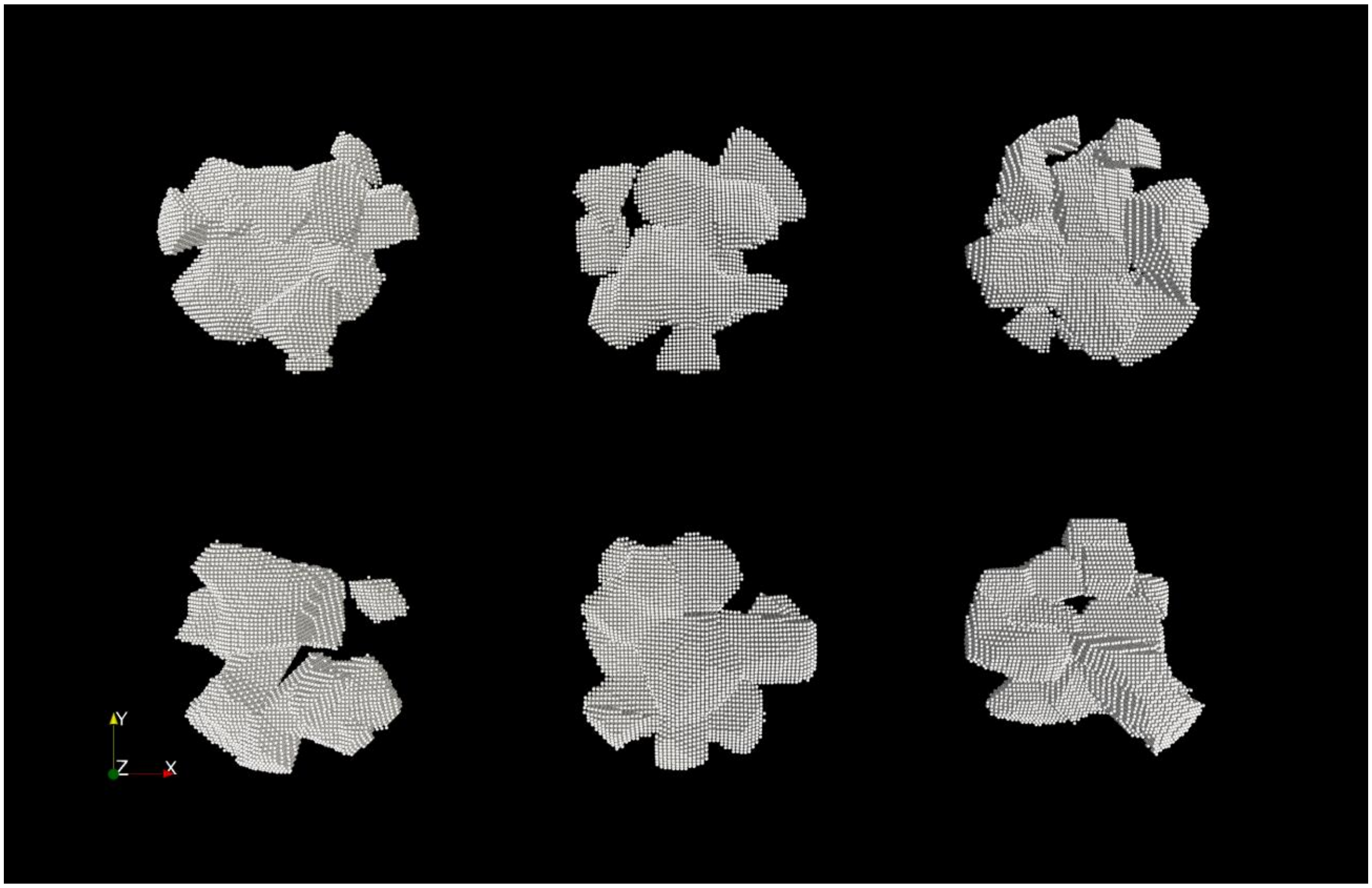

Figure 3: $\beta$-ratio vs. grain size calculated for the stellar properties of HR 4796A given in Table 2. Pure end-members are shown in 3A (top), while mixtures are shown in 3B (bottom). The percentages of silicate (Si), amorphous carbon (C), and $H_2O$ ice in 3B refer to proportion of solid, non-pore space.

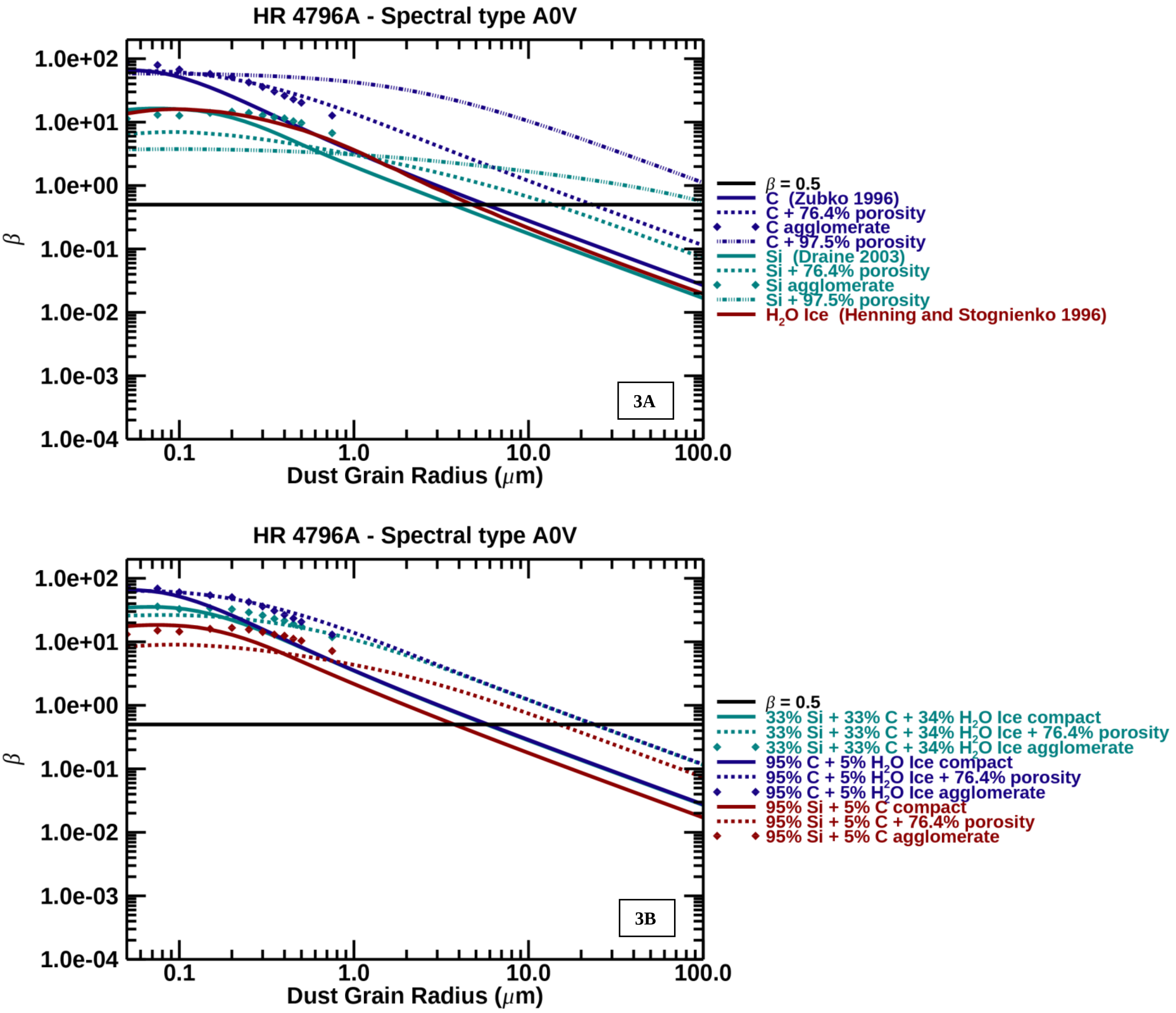

Figure 4: $\beta$-ratio vs. grain size calculated for the stellar properties of β Pictoris given in Table 2. Pure end-members are shown in 4A (top), while mixtures are shown in 4B (bottom). The percentages of silicate (Si), amorphous carbon (C), and $H_2O$ ice in 4B refer to proportion of solid, non-pore space.

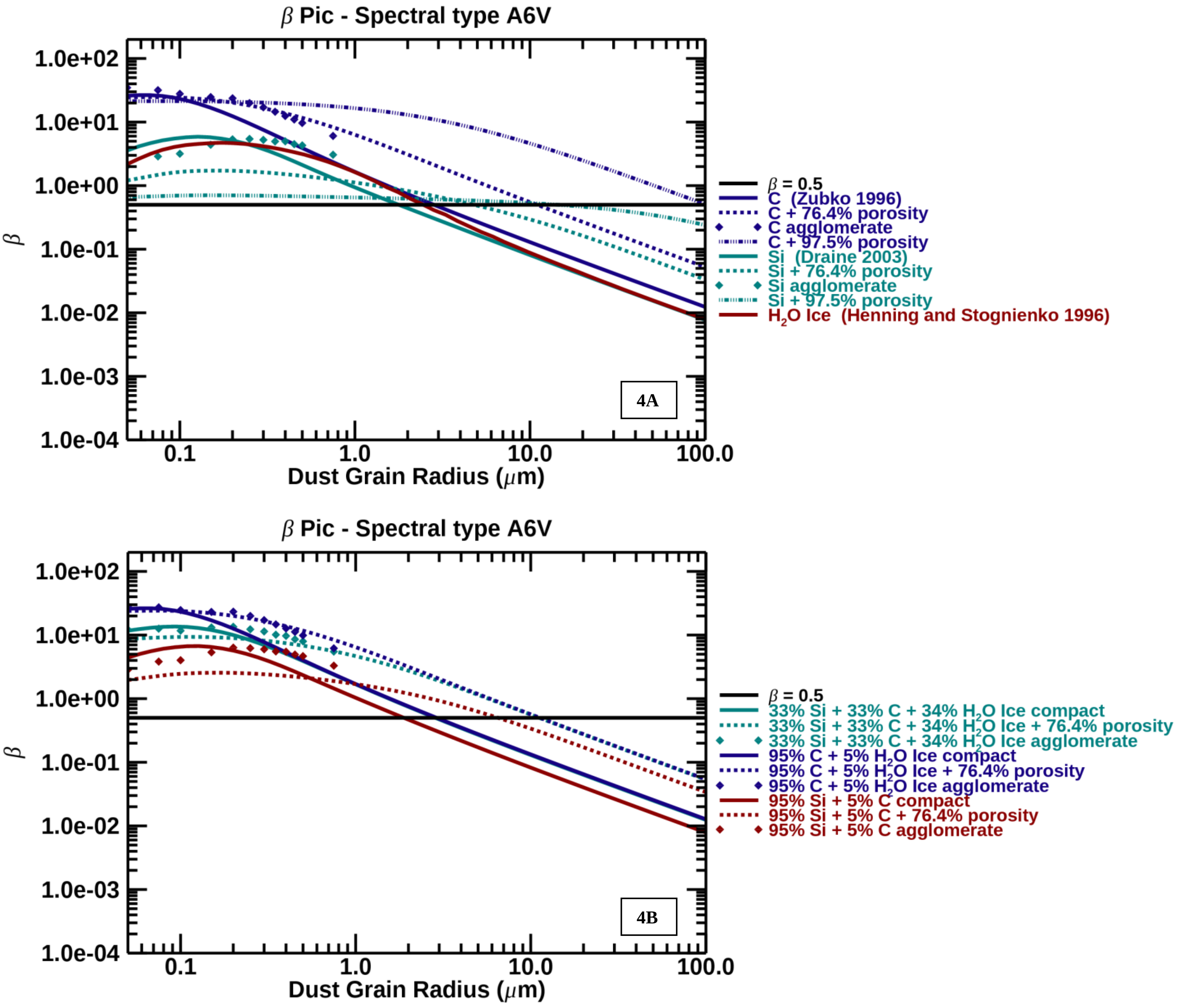

Figure 5: $\beta$-ratio vs. grain size calculated for the stellar properties of HD 32297 given in Table 2. Pure end-members are shown in 5A (top), while mixtures are shown in 5B (bottom). The percentages of silicate (Si), amorphous carbon (C), and $H_2O$ ice in 5B refer to proportion of solid, non-pore space.

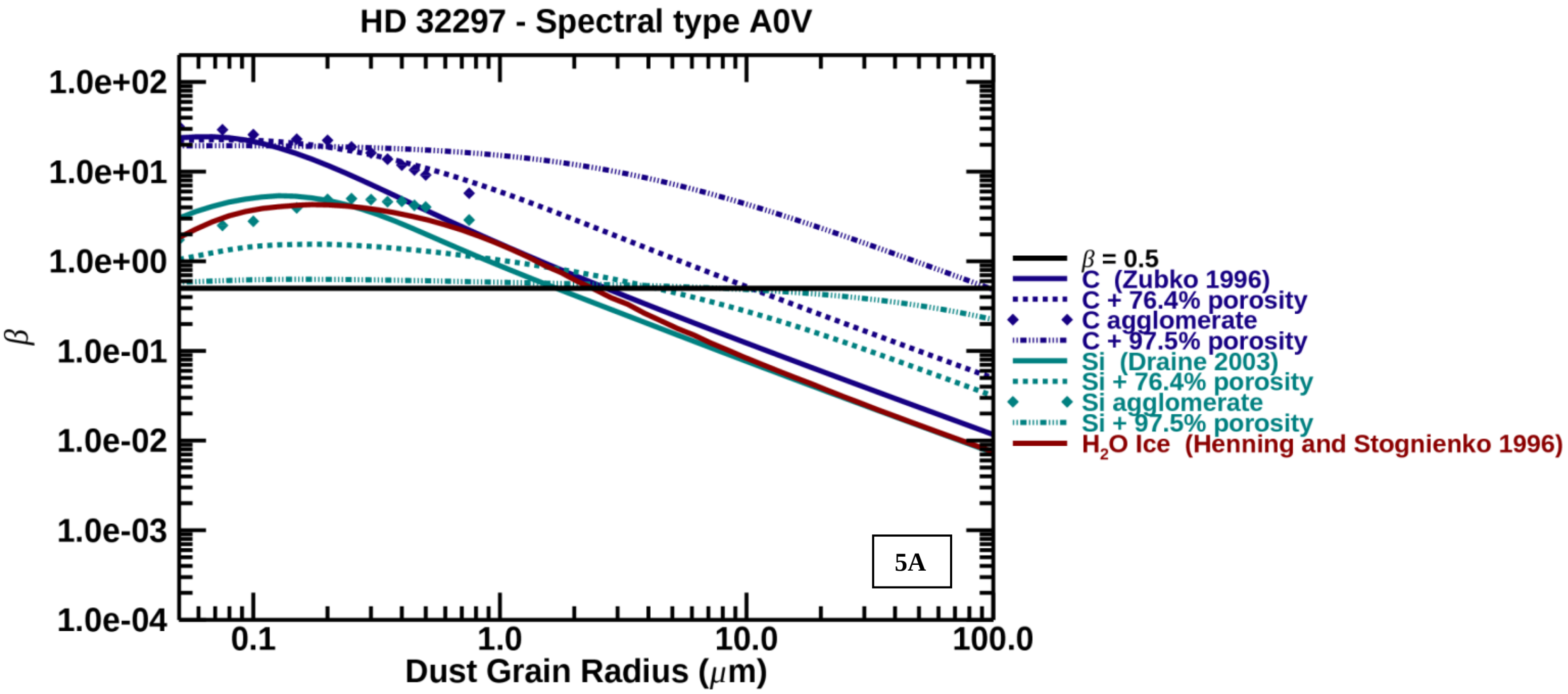

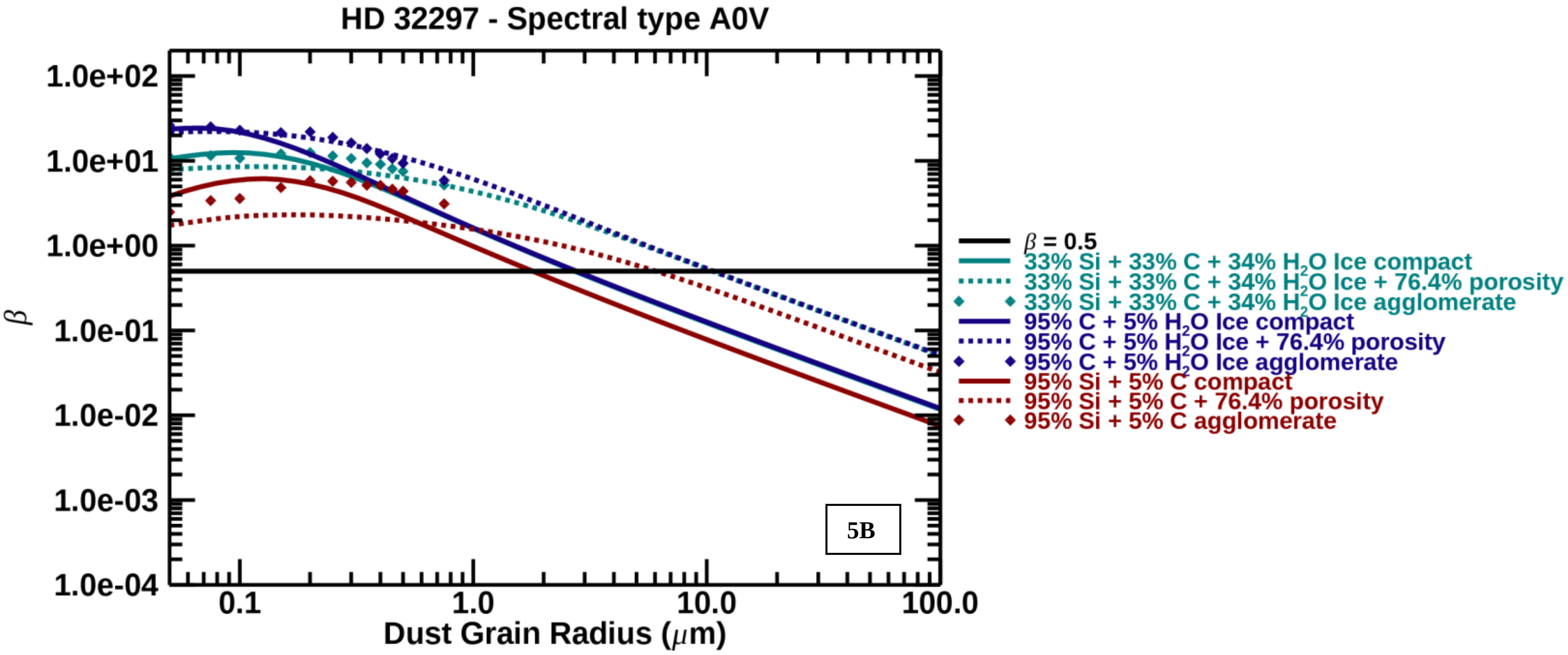

Figure 6: $\beta$-ratio vs. grain size calculated for the stellar properties of HD 181327 given in Table 2. Pure end-members are shown in 6A (top), while mixtures are shown in 6B (bottom). The percentages of silicate (Si), amorphous carbon (C), and $H_2O$ ice in 6B refer to proportion of solid, non-pore space.

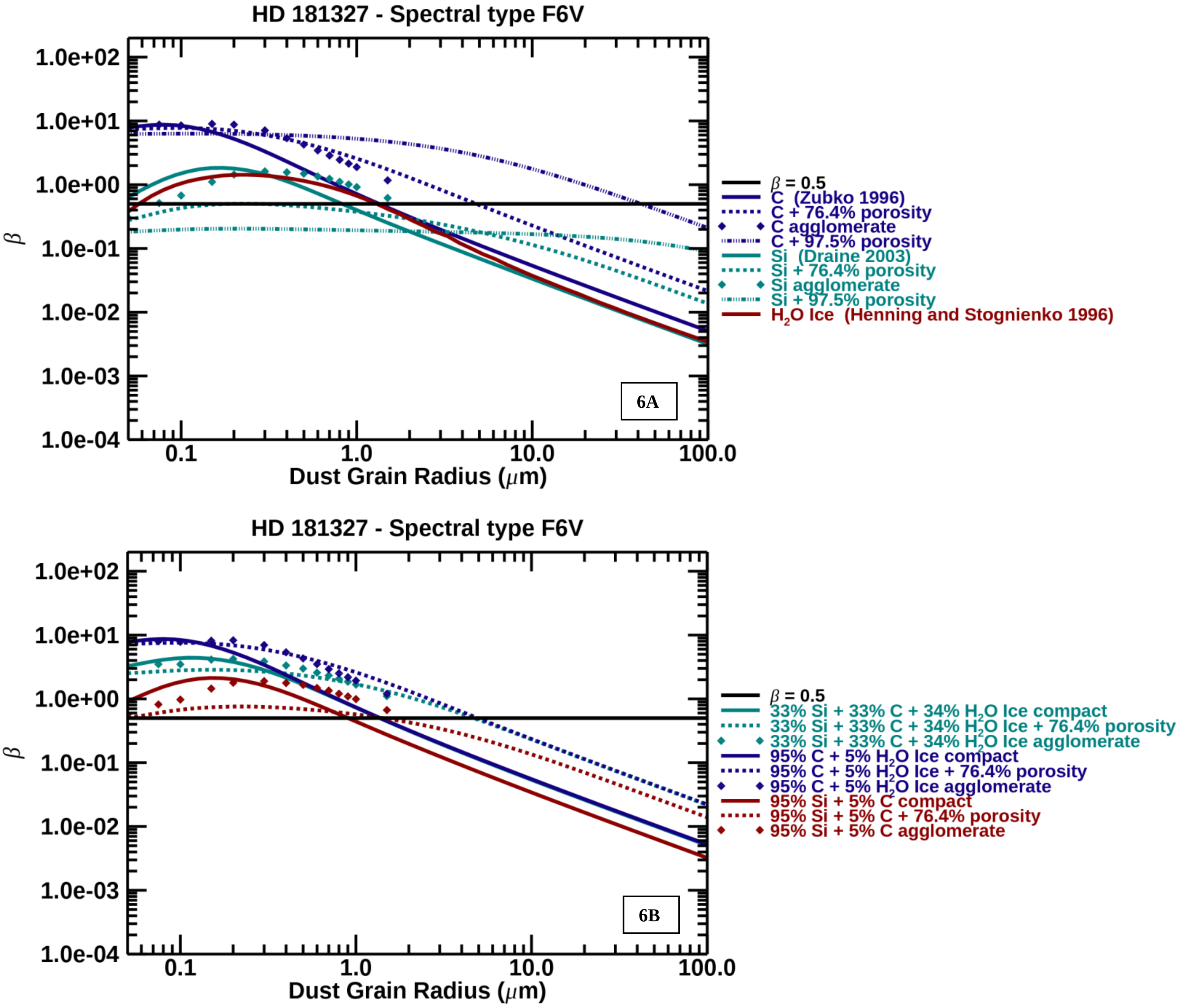

Figure 7: $\beta$-ratio vs. grain size calculated for the stellar properties of BD +20°307 given in Table 2. Pure end-members are shown in 7A (top), while mixtures are shown in 7B (bottom). The percentages of silicate (Si), amorphous carbon (C), and $H_2O$ ice in 7B refer to proportion of solid, non-pore space.

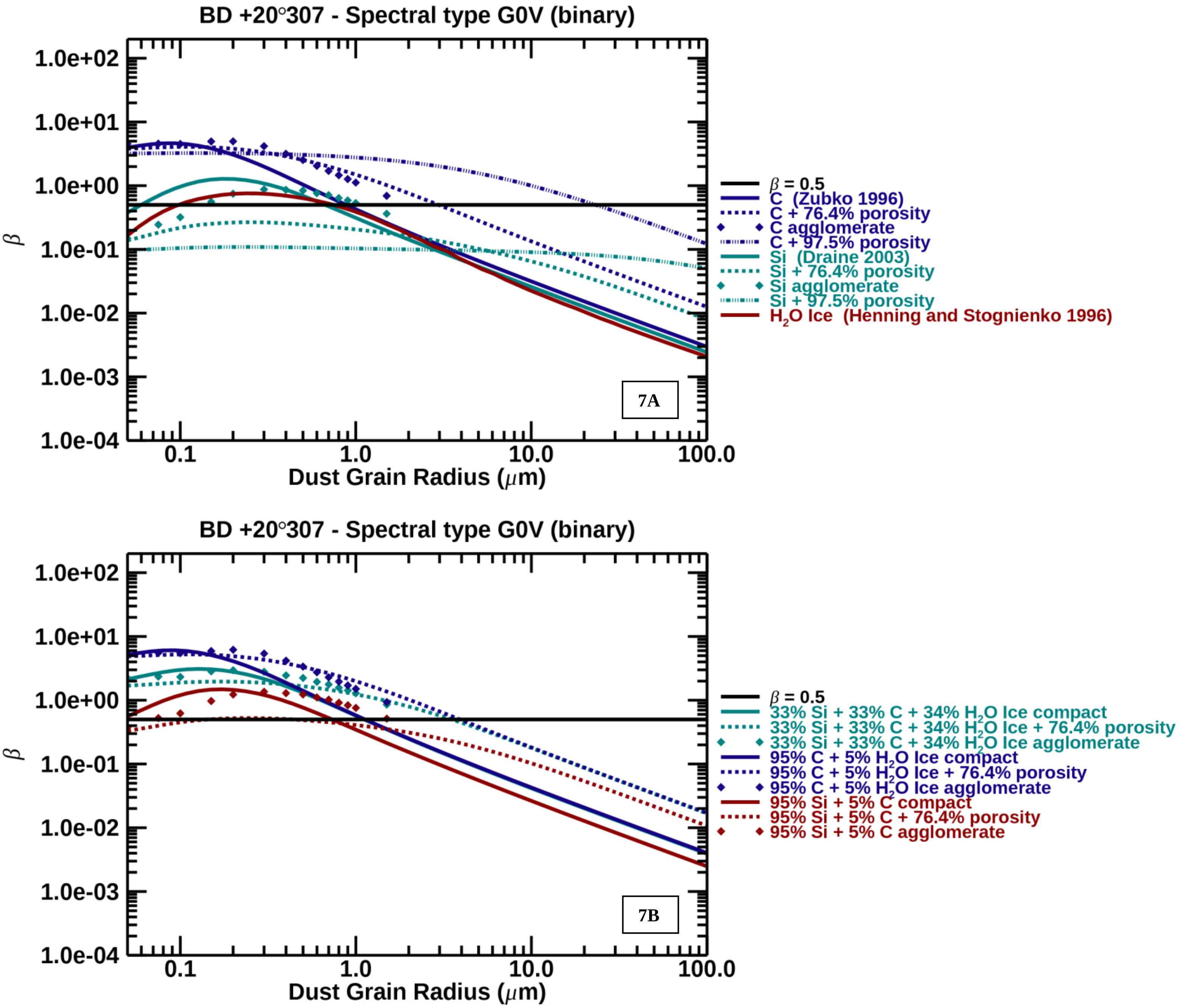

Figure 8: $\beta$-ratio vs. grain size calculated for the stellar properties of HD 61005 given in Table 2. Pure end-members are shown in 8A (top), while mixtures are shown in 8B (bottom). The percentages of silicate (Si), amorphous carbon (C), and $H_2O$ ice in 8B refer to proportion of solid, non-pore space.

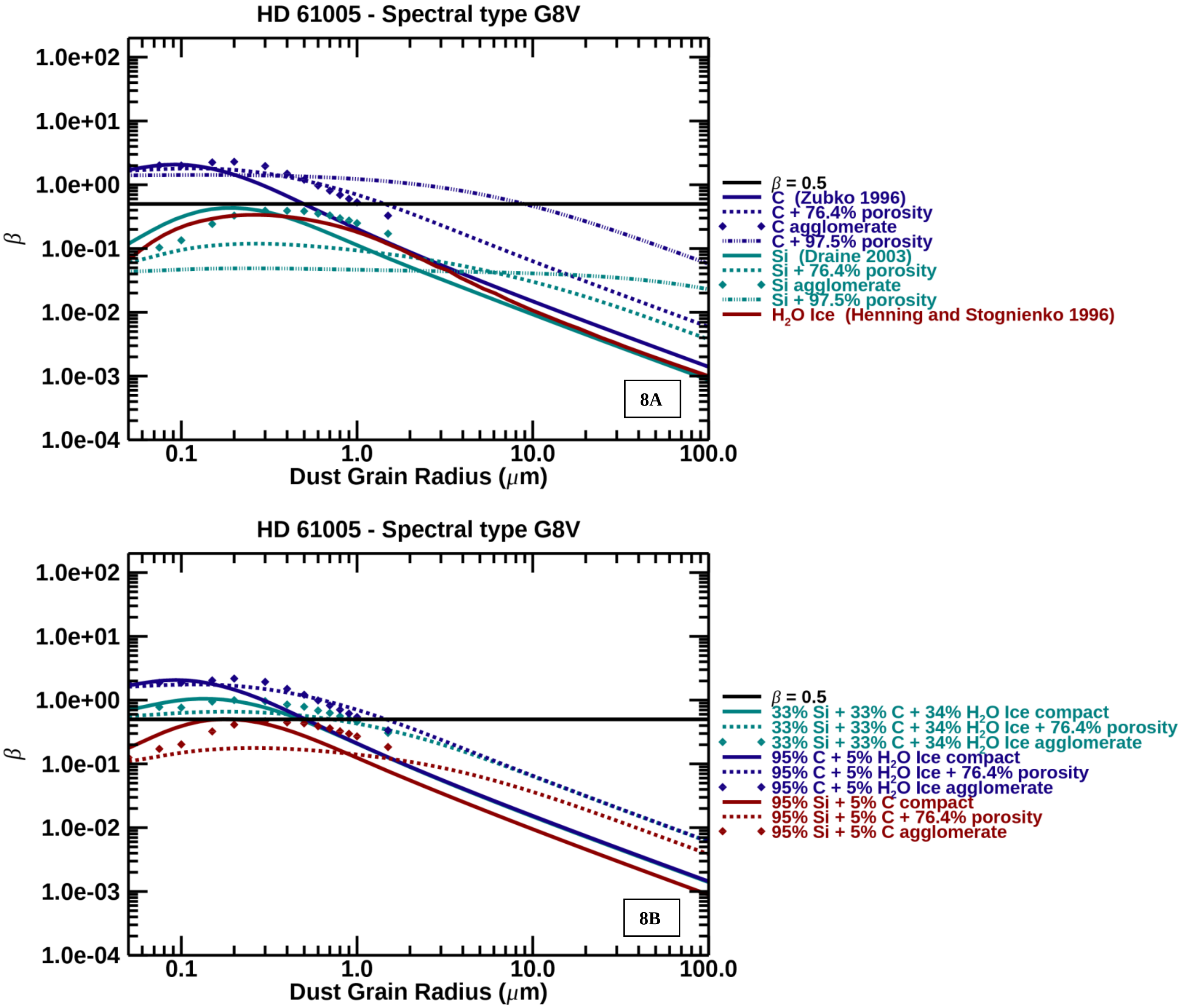

Figure 9: $\beta$-ratio vs. grain size calculated for the stellar properties of ε Eridani given in Table 2. Pure end-members are shown in 9A (top), while mixtures are shown in 9B (bottom). The percentages of silicate (Si), amorphous carbon (C), and $H_2O$ ice in 9B refer to proportion of solid, non-pore space.

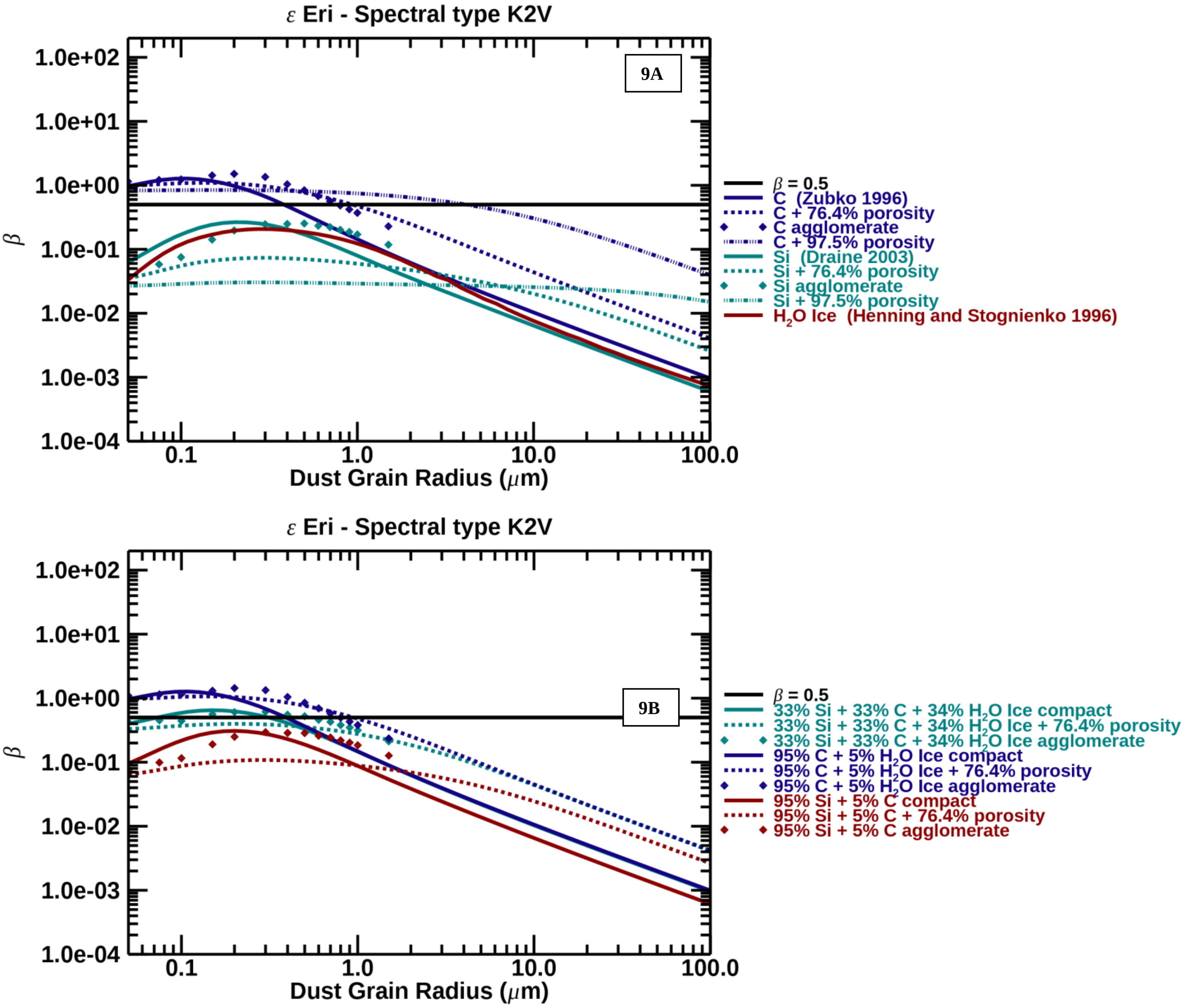

Figure 10: $\beta$-ratio vs. grain size calculated for the stellar properties of AU Microscopii given in Table 2. Pure end-members are shown in 10A (top), while mixtures are shown in 10B (bottom). The percentages of silicate (Si), amorphous carbon (C), and $H_2O$ ice in 10B refer to proportion of solid, non-pore space.

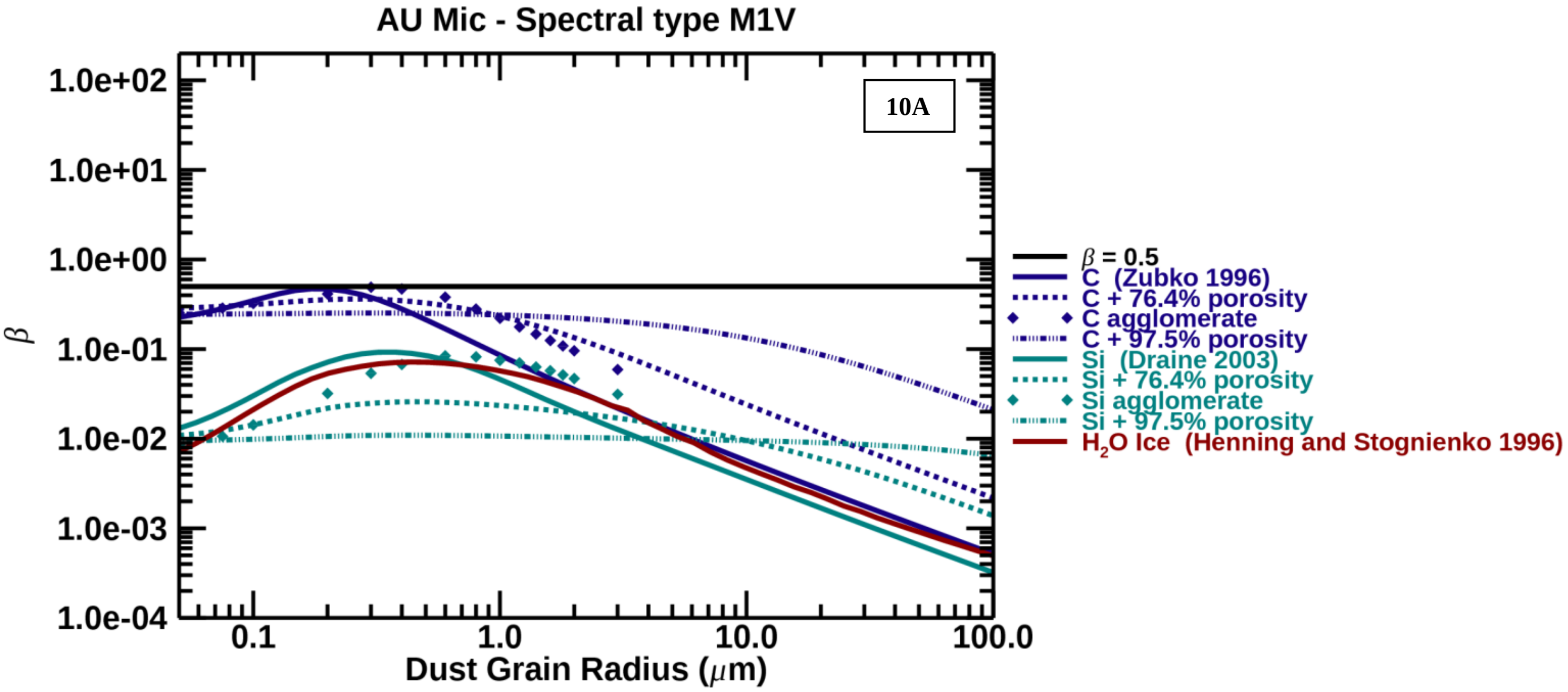

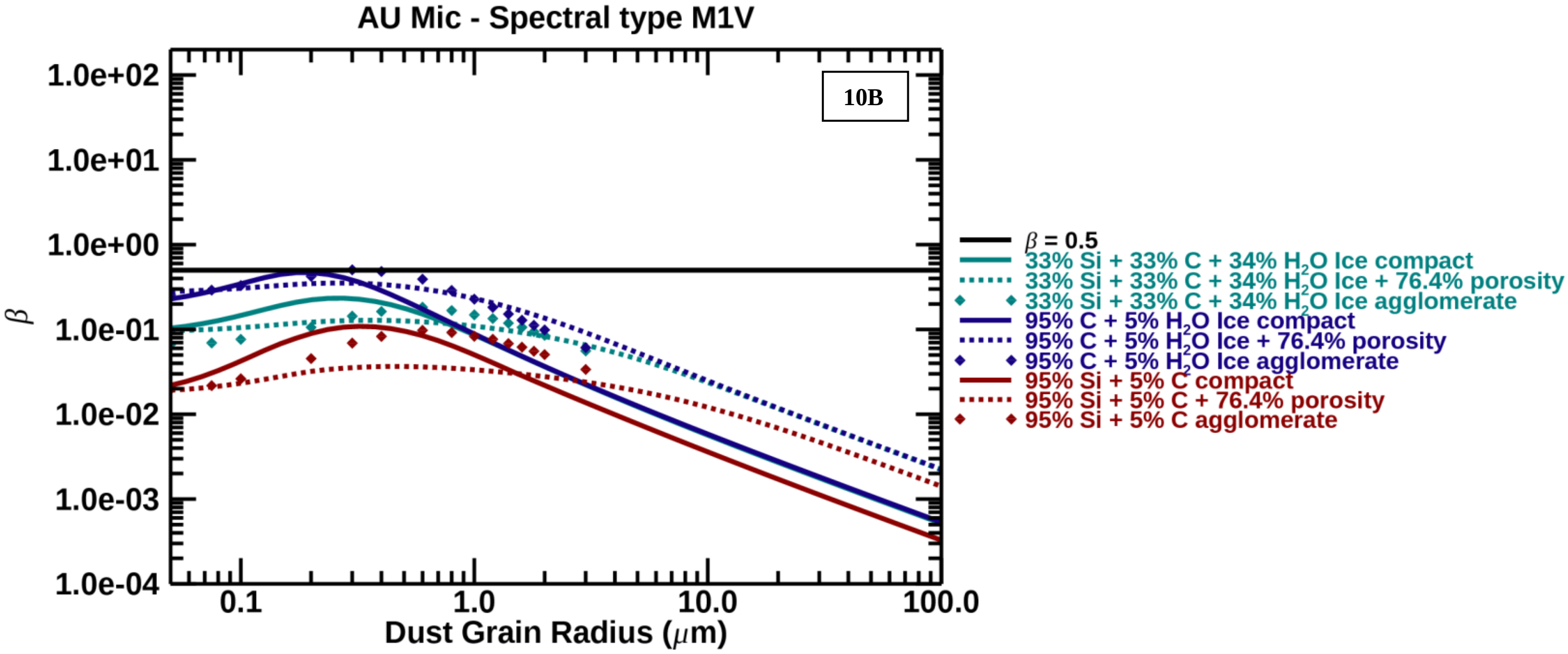

Figure 11: $\beta$-ratio vs. grain size calculated for the Sun. Pure end-members are shown in 11A (top), while mixtures are shown in 11B (bottom). The percentages of silicate (Si), amorphous carbon (C), and $H_2O$ ice in 11B refer to proportion of solid, non-pore space.

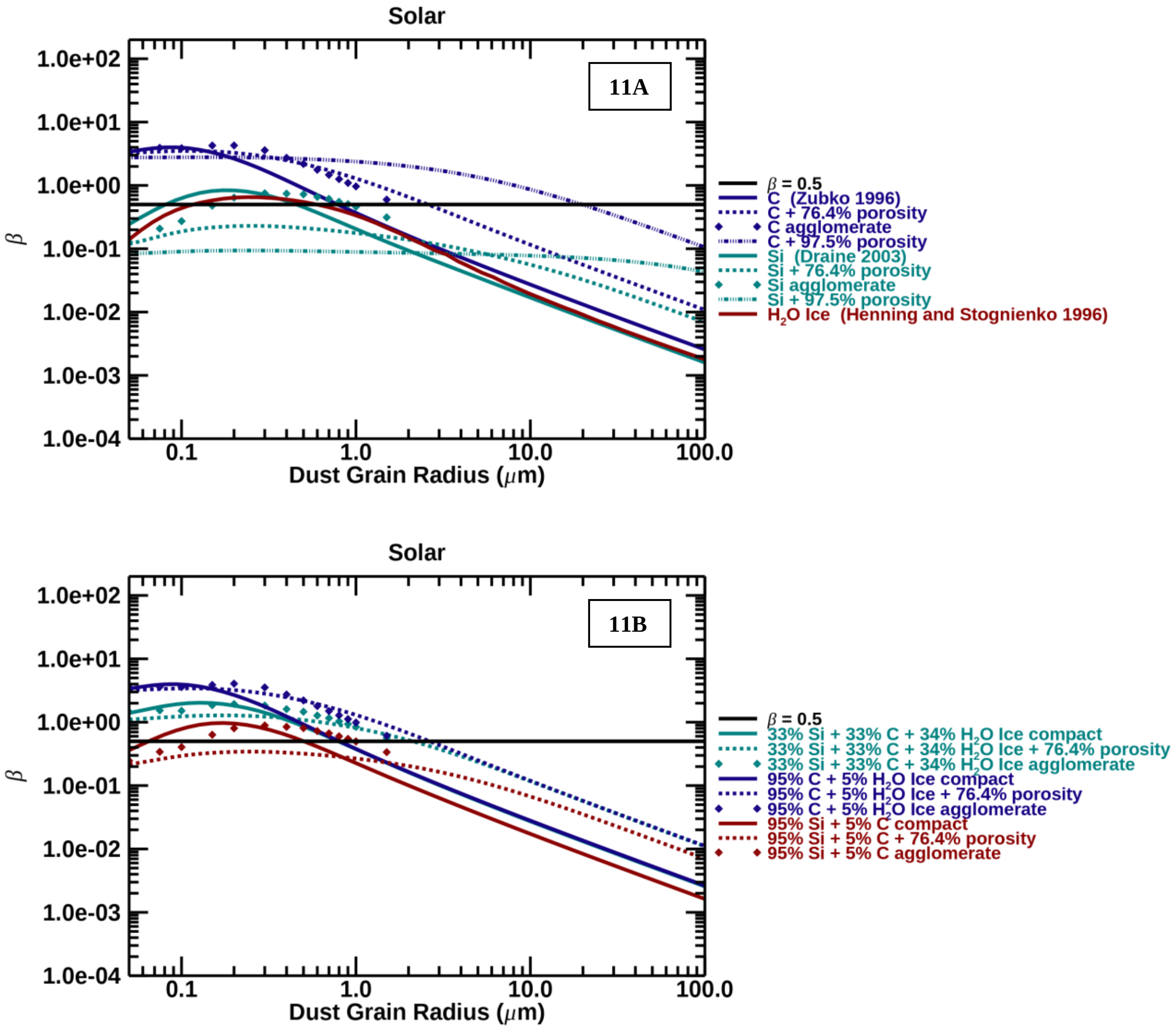

Table 1: Size parameters, size of the dipole lattice, and indices of refraction for the agglomerated debris particles in our database. The size parameter step is the increment between each of the size parameters included in the database for each column.

| | **Size parameter** | **0.1-0.9** | **1-16** | **17** | **18** | **19** | **20-26** | **28-32** | **32-44** | **48** |
|---|---|---|---|---|---|---|---|---|---|---|
| | Size parameter step | 0.1 | 1 | 1 | 1 | 1 | 2 | 2 | 2 | - |
| | Lattice diameter (# of dipoles) | 8 | 16-128 | 128 | 128 | 128 | 128-256 | 256 | 256 | 256 |
| **n** | **k** | | | | | | | | | |
| **1.1** | **0** | ✓ | ✓ | ✓ | ✓ | ✓ | ✓ | ✓ | ✓ | |
| **1.2** | **0** | ✓ | ✓ | ✓ | ✓ | ✓ | ✓ | ✓ | ✓ | |
| | **0.015** | ✓ | ✓ | ✓ | ✓ | ✓ | ✓ | ✓ | ✓ | |
| **1.313** | **0** | ✓ | ✓ | ✓ | ✓ | | ✓ | ✓ | ✓ | |
| | **0.02** | ✓ | ✓ | ✓ | ✓ | | ✓ | ✓ | ✓ | |
| | **0.05** | ✓ | ✓ | ✓ | ✓ | | ✓ | ✓ | ✓ | |
| | **0.1** | ✓ | ✓ | ✓ | ✓ | | ✓ | ✓ | ✓ | |
| **1.4** | **0** | ✓ | ✓ | | ✓ | | ✓ | ✓ | | |
| | **0.01** | ✓ | ✓ | | ✓ | | ✓ | ✓ | | |
| | **0.0175** | ✓ | ✓ | | ✓ | | ✓ | ✓ | | |
| | **0.02** | ✓ | ✓ | | ✓ | | ✓ | ✓ | | |
| | **0.05** | ✓ | ✓ | | ✓ | | ✓ | ✓ | | |
| | **0.1** | ✓ | ✓ | | ✓ | | ✓ | ✓ | | |
| | **1** | ✓ | ✓ | | ✓ | | ✓ | ✓ | | |
| **1.5** | **0** | ✓ | ✓ | | ✓ | | ✓ | ✓ | | |
| | **0.01** | ✓ | ✓ | | ✓ | | ✓ | ✓ | | |
| | **0.02** | ✓ | ✓ | | ✓ | | ✓ | ✓ | | |
| | **0.05** | ✓ | ✓ | | ✓ | | ✓ | ✓ | | |
| | **0.1** | ✓ | ✓ | | ✓ | | ✓ | ✓ | | |
| | **0.2** | ✓ | ✓ | | ✓ | | ✓ | ✓ | | |
| | **0.5** | ✓ | ✓ | | ✓ | | ✓ | ✓ | | |
| | **1** | ✓ | ✓ | | ✓ | | ✓ | ✓ | | |
| **1.6** | **0.0005** | ✓ | ✓ | | ✓ | | ✓ | ✓ | | |
| | **0.01** | ✓ | ✓ | | ✓ | | ✓ | ✓ | | |
| | **0.02** | ✓ | ✓ | | ✓ | | ✓ | ✓ | | ✓ |
| | **0.03** | ✓ | ✓ | | ✓ | | ✓ | ✓ | | |
| | **0.05** | ✓ | ✓ | | ✓ | | ✓ | ✓ | | |
| | **0.07** | ✓ | ✓ | | ✓ | | ✓ | ✓ | | |
| | **0.1** | ✓ | ✓ | | ✓ | | ✓ | ✓ | | |
| | **0.15** | ✓ | ✓ | | ✓ | | ✓ | ✓ | | |
| | **0.3** | ✓ | ✓ | | ✓ | | ✓ | ✓ | | |
| | **0.5** | ✓ | ✓ | | ✓ | | ✓ | ✓ | | |
| | **0.7** | ✓ | ✓ | | ✓ | | ✓ | ✓ | | |
| | **1** | ✓ | ✓ | | ✓ | | ✓ | ✓ | | |
| **1.7** | **0** | ✓ | ✓ | | ✓ | | ✓ | | | |
| | **0.02** | ✓ | ✓ | | ✓ | | ✓ | ✓ | | |
| | **0.05** | ✓ | ✓ | | ✓ | | ✓ | ✓ | | |
| | **0.1** | ✓ | ✓ | | ✓ | | ✓ | ✓ | | |
| **1.758** | **0.0844** | ✓ | ✓ | | ✓ | | ✓ | ✓ | | |
| **1.855** | **0.45** | ✓ | ✓ | | ✓ | | ✓ | ✓ | | |
| **1.957** | **0.341** | ✓ | ✓ | | ✓ | | ✓ | ✓ | | |
| **2.43** | **0.59** | ✓ | ✓ | | ✓ | | ✓ | ✓ | | ✓ |

Table 2: Stellar properties used to calculate the stellar spectrum and $\beta_{pr}$.

| **Host star** | **Alternate name(s)** | **Spectral type** | **L* (L⊙)** | **T* (K)** | **M* (M⊙)** | **Age (Myr)** |
|---|---|---|---|---|---|---|
| HR 4796† | HD 109537 | A0V/M2.5V [a] | 23 [b] | 9378 [c] | 2.18 [b] | 8 [d] |
| β Pic | HD 39060, HR 2020 | A6V [a] | 8.5* | 8000* | 1.75 [e] | <10 [e] |
| HD 32297 | BD +07°777 | A0 [a] | 7.8* | 7750* | 1.7 [f] | <30 [g] |
| HD 181327 | | F6V [a] | 2.7* | 6250* | 1.36 [h] | 12 [i] |
| BD +20°307† | | G0V/G0V [a] | 2.9* | 5750* | 0.9/0.9 [j] | 300 [k] |
| HD 61005 | BD +31°4778 | G8V [a] | 0.6 [*, l] | 5500 [*, l] | 1.1 [m] | 40 [l] |
| ε Eri | HD 22049, HR 1084 | K2V [a] | 0.3* | 5000* | 0.8 [n] | 400-800 [o] |
| AU Mic | HD 197481 | M1Ve [a] | 0.1* | 3300* | 0.5 [p] | 20 [p] |
| Sun | | G2V | 1 | 5778 | 1 | 4603 |

† binary system
* this work
a. simbad.u-strasbg.fr/simbad/
b. Gerbaldi et al. 1999
c. Saffe et al. 2008
d. Schneider 2009
e. Crifo et al. 1997
f. Debs et al. 2009
g. Kalas et al. 2005
h. Lebreton et al. 2017
i. Mamajek et al. 2004
j. Dotter et al. 2008
k. Song et al. 2005
l. Desidera et al. 2011
m. Olofsson et al. 2018
n. Greaves et al 1998
o. Mamajek & Hillenbrand 2008
p. Metchev, Eisner & Hillenbrand 2013

Table 3: Dust grain blowout sizes in microns inferred from fits to the β-ratio plots presented in Figures 3-10 for different grain models of dust surrounding stars with properties given in Table 2. The agglomerate calculations were extrapolated according to the linear fits given in table 4 where necessary.

| | **Disk** | **HR 4796** | **β Pic** | **HD 32297** | **HD 181327** | **BD +20°307** | **Sun** | **HD 61005** | **ε Eri** | **AU Mic** |
|---|---|---|---|---|---|---|---|---|---|---|
| Composition | Grain model type | | | | | | | | | |
| Amorphous carbon | Mie compact | 5.6 | 2.8 | 2.7 | 1.3 | 1.1 | 0.8 | 0.5 | 0.4 | * |
| | Mie 76.4% porous | 23 | 11 | 11 | 4.9 | 3.8 | 2.6 | 1.4 | 0.9 | * |
| | Agglomerate 75% porous | 11 | 6.1 | 5.8 | 3.1 | 2.5 | 1.7 | 1.1 | 0.8 | 0.2 |
| | Mie 97.5% porous | 220 | 100 | 95 | 42 | 32 | 20 | 8.8 | 4.0 | * |
| Astronomical silicate | Mie compact | 3.5 | 1.8 | 1.8 | 0.8 | 0.6 | 0.4 | * | * | * |
| | Mie 76.4% porous | 13 | 4.7 | 4.2 | * | * | * | * | * | * |
| | Agglomerate 76.4% porous | 11 | 4.9 | 4.6 | 1.9 | 1.4 | 0.9 | * | * | * |
| | Mie 97.5% porous | 120 | 13 | 7.9 | * | * | * | * | * | * |
| Water ice | Mie compact | 4.8 | 2.5 | 2.4 | 1.2 | 0.9 | 0.6 | * | * | * |
| 33% Si + 33% C + 34% Ice | Mie compact | 5.6 | 2.8 | 2.9 | 1.4 | 1.2 | 0.8 | 0.4 | 0.3 | * |
| | Mie 76.4% porous | 23 | 11 | 11 | 4.7 | 3.5 | 2.0 | 0.7 | * | * |
| | Agglomerate 76.4% porous | 18 | 8.3 | 7.8 | 3.4 | 2.6 | 1.7 | 0.9 | 0.5 | * |
| 95% C + 5% Ice | Mie compact | 5.8 | 2.9 | 2.7 | 1.4 | 1.1 | 0.8 | 0.5 | 0.4 | * |
| | Mie 76.4% porous | 24 | 11 | 11 | 5.1 | 3.9 | 2.7 | 1.5 | 0.9 | * |
| | Agglomerate 76.4% porous | 11 | 6.2 | 5.9 | 3.2 | 2.6 | 1.8 | 1.1 | 0.8 | * |
| 95% Si + 5% C | Mie compact | 3.7 | 1.9 | 1.9 | 0.9 | 0.7 | 0.5 | 0.2 | * | * |
| | Mie 76.4% porous | 15 | 6.4 | 6.0 | 1.5 | 0.5 | * | * | * | * |
| | Agglomerate 76.4% porous | 13 | 5.6 | 5.3 | 2.1 | 1.6 | 1.0 | * | * | * |

* Minimum grain size not constrained by radiation pressure.

Table 4: Six parameter Moffat distribution fits (C0-C5) to agglomerated debris β-ratio versus grain radius ($a$ in μm) curves for each composition in Figures 3-10. These are only valid for the grain sizes shown in Figures 3-10 (0.05 μm – 100 μm). The functional form of the Moffat fit is $\log_{10}\beta = \frac{C0}{(u^2+1)^{C3}} + C4 + C5\log_{10} a$, where $u = \frac{\log_{10} a - C1}{C2}$. Linear fits to the last four points ($\log_{10}\beta = L0 + L1\log_{10} a$) that were used to extrapolate the agglomerate calculations when necessary to find $a_{min}$.

| **Host star** | | **Amorphous carbon** | **Astronomical silicate** | **33% Amorphous carbon 33% Astronomical silicate 34% Water ice** | **95% Amorphous carbon 5% Water ice** | **95% Astronomical silicate 5% Amorphous carbon** |
|---|---|---|---|---|---|---|
| HR 4796 | C0 | 1.09E+04 | 1.65E+04 | 2.22E+00 | 1.21E+04 | 1.56E+04 |
| | C1 | -6.46E-01 | -5.30E-01 | -6.09E-01 | -6.31E-01 | -5.67E-01 |
| | C2 | 3.52E-01 | 5.65E-01 | 4.08E-01 | 3.22E-01 | 5.10E-01 |
| | C3 | 1.29E-05 | 2.02E-05 | 8.70E-02 | 1.19E-05 | 1.84E-05 |
| | C4 | -1.09E+04 | -1.65E+04 | -1.04E+00 | -1.21E+04 | -1.56E+04 |
| | C5 | -7.63E-01 | -3.77E-01 | -4.78E-01 | -7.03E-01 | -3.64E-01 |
| | L0 | 9.56E-01 | 7.07E-01 | 9.49E-01 | 9.65E-01 | 7.39E-01 |
| | L1 | -1.20E+00 | -9.67E-01 | -9.98E-01 | -1.20E+00 | -9.34E-01 |
| β Pic | C0 | 1.24E+04 | 1.78E+04 | 3.48E-01 | 1.36E+04 | 1.83E+04 |
| | C1 | -6.17E-01 | -5.62E-01 | -5.89E-01 | -6.12E-01 | -5.75E-01 |
| | C2 | 2.82E-01 | 4.66E-01 | 8.29E-01 | 2.68E-01 | 4.30E-01 |
| | C3 | 1.05E-05 | 2.01E-05 | 2.60E+00 | 1.04E-05 | 1.67E-05 |
| | C4 | -1.24E+04 | -1.78E+04 | 5.25E-01 | -1.36E+04 | -1.83E+04 |
| | C5 | -7.09E-01 | -4.37E-02 | -3.70E-01 | -6.36E-01 | -1.05E-01 |
| | L0 | 6.33E-01 | 3.65E-01 | 6.15E-01 | 6.43E-01 | 4.02E-01 |
| | L1 | -1.20E+00 | -9.67E-01 | -9.98E-01 | -1.20E+00 | -9.34E-01 |
| HD 32297 | C0 | 1.26E+04 | 1.91E+04 | 3.41E-01 | 1.38E+04 | 1.81E+04 |
| | C1 | -6.14E-01 | -5.60E-01 | -5.85E-01 | -6.10E-01 | -5.71E-01 |
| | C2 | 2.75E-01 | 4.63E-01 | 8.77E-01 | 2.62E-01 | 4.24E-01 |
| | C3 | 1.03E-05 | 1.92E-05 | 2.98E+00 | 1.03E-05 | 1.68E-05 |
| | C4 | -1.26E+04 | -1.91E+04 | 5.08E-01 | -1.38E+04 | -1.81E+04 |
| | C5 | -7.00E-01 | -1.36E-02 | -3.59E-01 | -6.27E-01 | -8.38E-02 |
| | L0 | 6.11E-01 | 3.40E-01 | 5.90E-01 | 6.20E-01 | 3.76E-01 |
| | L1 | -1.20E+00 | -9.67E-01 | -9.98E-01 | -1.20E+00 | -9.34E-01 |
| HD 181327 | C0 | 1.61E+04 | -2.63E+00 | 5.20E-01 | 1.63E+04 | -2.08E+00 |
| | C1 | -5.75E-01 | -1.73E+00 | -4.79E-01 | -5.49E-01 | -1.64E+00 |
| | C2 | 3.92E-01 | 3.19E+02 | 4.07E+02 | 3.93E-01 | 3.56E+02 |
| | C3 | 1.51E-05 | 1.23E+05 | 2.97E+05 | 1.52E-05 | 1.77E+05 |
| | C4 | -1.61E+04 | 3.42E-02 | -1.26E-01 | -1.63E+04 | 4.54E-02 |
| | C5 | -5.81E-01 | -1.19E+00 | -3.74E-01 | -5.60E-01 | -1.13E+00 |
| | L0 | 2.76E-01 | -4.23E-02 | 2.20E-01 | 2.86E-01 | -8.34E-03 |
| | L1 | -1.18E+00 | -9.14E-01 | -9.81E-01 | -1.17E+00 | -9.26E-01 |
| BD +20°307 | C0 | 1.68E+04 | -2.46E+00 | -1.53E+00 | 1.70E+04 | -1.99E+00 |
| | C1 | -5.62E-01 | -1.64E+00 | -1.52E+00 | -5.36E-01 | -1.57E+00 |
| | C2 | 3.82E-01 | 1.48E+02 | 1.97E+00 | 3.78E-01 | 2.61E+02 |
| | C3 | 1.51E-05 | 2.99E+04 | 6.00E+00 | 1.47E-05 | 1.04E+05 |
| | C4 | -1.68E+04 | -2.11E-01 | 8.07E-02 | -1.70E+04 | -1.93E-01 |
| | C5 | -5.45E-01 | -1.14E+00 | -1.20E+00 | -5.30E-01 | -1.10E+00 |
| | L0 | 1.68E-01 | -1.59E-01 | 1.05E-01 | 1.79E-01 | -1.25E-01 |
| | L1 | -1.18E+00 | -8.88E-01 | -9.62E-01 | -1.18E+00 | -9.03E-01 |
| HD 61005 | C0 | 1.72E+04 | -8.50E+03 | -1.58E+00 | 1.75E+04 | -1.97E+00 |
| | C1 | -5.56E-01 | -4.09E+00 | -1.52E+00 | -5.30E-01 | -1.54E+00 |
| | C2 | 3.78E-01 | 8.86E-04 | 2.14E+00 | 3.71E-01 | 1.97E+02 |
| | C3 | 1.51E-05 | 5.32E-04 | 6.80E+00 | 1.46E-05 | 6.13E+04 |
| | C4 | -1.72E+04 | 8.42E+03 | -2.42E-01 | -1.75E+04 | -5.22E-01 |
| | C5 | -5.24E-01 | -2.48E+00 | -1.21E+00 | -5.13E-01 | -1.09E+00 |
| | L0 | -2.75E-01 | -6.08E-01 | -3.43E-01 | -2.65E-01 | -5.74E-01 |
| | L1 | -1.19E+00 | -8.72E-01 | -9.50E-01 | -1.18E+00 | -8.90E-01 |
| ε Eri | C0 | -1.24E+00 | -2.53E+04 | -1.68E+00 | -1.32E+00 | -1.98E+00 |
| | C1 | -1.43E+00 | -4.52E-01 | -1.53E+00 | -1.45E+00 | -1.51E+00 |
| | C2 | 6.09E+02 | 5.98E-01 | 3.02E+00 | 6.34E+02 | 9.25E+00 |
| | C3 | 9.65E+05 | -2.47E-05 | 1.24E+01 | 9.33E+05 | 1.38E+02 |
| | C4 | -4.22E-01 | 2.53E+04 | -4.03E-01 | -4.06E-01 | -6.81E-01 |
| | C5 | -1.29E+00 | 2.06E-01 | -1.22E+00 | -1.32E+00 | -1.09E+00 |
| | L0 | -4.31E-01 | -7.75E-01 | -5.07E-01 | -4.20E-01 | -7.40E-01 |
| | L1 | -1.19E+00 | -8.39E-01 | -9.25E-01 | -1.18E+00 | -8.60E-01 |
| AU Mic | C0 | -4.84E-01 | -2.98E+04 | -2.45E+04 | 2.23E+04 | -2.64E+04 |
| | C1 | -4.64E-01 | -1.82E-01 | -1.39E-01 | -3.53E-01 | -1.64E-01 |
| | C2 | 2.75E-01 | 5.16E-01 | 4.39E-01 | 4.69E-01 | 4.55E-01 |
| | C3 | -3.16E-01 | -1.75E-05 | -1.37E-05 | 1.75E-05 | -1.48E-05 |
| | C4 | 1.85E-04 | 2.98E+04 | 2.45E+04 | -2.23E+04 | 2.64E+04 |
| | C5 | -4.11E-01 | 1.21E-01 | -2.10E-01 | -3.95E-01 | -3.52E-02 |
| | L0 | -6.58E-01 | -1.05E+00 | -7.73E-01 | -6.45E-01 | -1.02E+00 |
| | L1 | -1.20E+00 | -9.67E-01 | -9.98E-01 | -1.20E+00 | -9.34E-01 |